\documentclass[sigconf]{acmart}

\usepackage{graphicx}   
\usepackage{booktabs}   
\usepackage{amsmath}    
\usepackage{bm}         
 
\usepackage{amssymb}    
\usepackage{accents}       
\usepackage{multirow,makecell}  
\usepackage{subfigure}  
\usepackage{CJK}        
\usepackage{epstopdf}
\usepackage{lineno}     
\usepackage{bm}         

\AtBeginDocument{%
  \providecommand\BibTeX{{%
    \normalfont B\kern-0.5em{\scshape i\kern-0.25em b}\kern-0.8em\TeX}}}

\copyrightyear{2021} 
\acmYear{2021} 
\setcopyright{acmcopyright}
\acmConference[CIKM '21]{Proceedings of the 30th ACM International Conference on Information and Knowledge Management}{November 1--5, 2021}{Virtual Event, QLD, Australia}
\acmBooktitle{Proceedings of the 30th ACM International Conference on Information and Knowledge Management (CIKM '21), November 1--5, 2021, Virtual Event, QLD, Australia}
\acmPrice{15.00}
\acmDOI{10.1145/3459637.3482189}
\acmISBN{978-1-4503-8446-9/21/11}

\begin{document}
\fancyhead{}

\title[ST-PIL: Spatial-Temporal Periodic Interest Learning for Next Point-of-Interest Recommendation]{ST-PIL: Spatial-Temporal Periodic Interest Learning for Next Point-of-Interest Recommendation}



\author[Q. Cui, C. Zhang, Y. Zhang, J. Wang, M. Chen]{Qiang Cui, Chenrui Zhang, Yafeng Zhang, Jinpeng Wang\textsuperscript{$\ast$}, Mingchen Cai} 

\makeatletter
\def\authornotetext#1{
  \if@ACM@anonymous\else
    \g@addto@macro\@authornotes{%
    \stepcounter{footnote}\footnotetext{#1}}%
\fi}
\makeatother

\authornotetext{Jinpeng Wang is the corresponding author.}

\affiliation{
  \institution{Meituan, Beijing, China}
  \country{}
  }
\email{{cuiqiang04, zhangchenrui02, zhangyafeng, wangjinpeng04,caimingchen}@meituan.com}

\def\authors{Qiang Cui, Chenrui Zhang, Yafeng Zhang, Jinpeng Wang, Mingchen Cai}



\renewcommand{\shortauthors}{Cui, et al.}





\begin{abstract}
  Point-of-Interest (POI) recommendation is an important task in location-based social networks. It facilitates the relation modeling between users and locations. Recently, researchers recommend POIs by long- and short-term interests and achieve success. 
  However, they fail to well capture the periodic interest. People tend to visit similar places at similar times or in similar areas. Existing models try to acquire such kind of periodicity by user's mobility status or time slot, which limits the performance of periodic interest. 
  To this end, we propose to learn spatial-temporal periodic interest. Specifically, in the long-term module, we learn the temporal periodic interest of daily granularity, then utilize intra-level attention to form long-term interest.
  In the short-term module, we construct various short-term sequences to acquire the spatial-temporal periodic interest of hourly, areal, and hourly-areal granularities, respectively. Finally, we apply inter-level attention to automatically integrate multiple interests. 
  Experiments on two real-world datasets demonstrate the state-of-the-art performance of our method. 
\end{abstract}



\maketitle

\section{Introduction}
  Location-based Social Networks (LBSNs), such as Foursquare, and Yelp, enable users to share check-in experiences and opinions on Point-of-Interests (POIs). As one of the core services of LBSNs, POI recommendation \cite{islam2020survey} focuses on utilizing user's check-ins, location-based information, contextual information, and social relationships to recommend POIs. This task is beneficial not only to the improvement of user experiences and social networking services but also to the advertising between potential locations and users. 

  A large number of methods are proposed to solve the next POI recommendation and many researchers recently combine long- and short-term interests to boost performance \cite{manotumruksa2017deep,feng2018deepmove,zhao2019go,wu2019long,wu2020personalized,sun2020go}.
  The long-term interest is usually acquired with all his/her historical check-ins, while the short-term interest is modeled with recent check-ins \cite{zhao2019go}. 
  These interests are usually encoded by feeding the user's check-ins to an encoder network, e.g., a long short-term memory (LSTM) \cite{hochreiter1997long} to capture transitional regularities. The way of organizing inputs as successive check-ins works well as users' recent check-ins are more likely to impact his/her next check-in \cite{manotumruksa2017deep,cui2019distance2pre,wu2020personalized,sun2020go}.
  
  However, in the case when the user's check-ins are abundant, it is difficult for the model to encode all POIs.
  Moreover, the location preference of users is generally periodic \cite{zhao2016survey}. As a result, only a small subset of historical check-ins are highly related to the user’s next visit. Hence, the successive check-ins would weaken the signals of the related parts.
  Only a few methods attempt to get out of this limitation by modeling periodicity. For example, DeepMove \cite{feng2018deepmove} applies the current mobility status to capture periodicity from all check-ins, LSTPM \cite{sun2020go} uses time slot to compute temporal similarity to obtain time-weighted trajectory representation, but they still suffer from the limited capability of learning periodic interest.

  The context of check-in, e.g., spatial or temporal information when a user is visiting the POI recommendation system, can provide some hints to the way of modeling periodic interest.
  Firstly, user activity is constrained by time and exhibits daily and hourly patterns. For instance, some users probably visit different places for vacation only during leisure time, e.g., weekends. Some other users would go to restaurants in the evening. 
  Secondly, user behavior shows an areal pattern when it comes to different areas. Such as some users visit historical sites in the scenic area but play chess in the entertainment area. 
  Thirdly, users will visit the same POI at a specific time, which refers to an hourly-areal pattern. For example, many people go to work at the company at 8 o'clock. 
  Previous methods like DeepMove \cite{feng2018deepmove} and LSTPM \cite{sun2020go} do not make full use of the spatial-temporal information when learning periodicity.

  Motivated by the above analysis, we propose a \textbf{S}patial-\textbf{T}emporal \textbf{P}eriodic \textbf{I}nterest \textbf{L}earning network (\textbf{ST-PIL}) for next POI recommendation. Specifically, ST-PIL adopts the structure of long- and short-term interest learning, but fully utilizes spatial-temporal context to retrieve related parts from historical check-ins for periodic interest. 
  For the long-term module, we split a user's all check-ins into different sequences, according to temporal context called day of week. Then we obtain the daily pattern by average pooling and acquire long-term interest by intra-level attention. This captures the user's temporal mobility of daily granularity.
  For the short-term module, we obtain various sequences by contexts called time slot and geohash\footnote{\url{https://en.wikipedia.org/wiki/Geohash}}. Correspondingly, LSTMs are deployed to obtain short-term interests. This module acquires periodicities of hourly, areal, and hourly-areal granularities. 
  For overall representation, inter-level attention and weighted concatenation are used to adaptively combine all interests. A multilayer perceptron is used to obtain the final probability for a candidate POI. 
  The main contributions are summarized as follows:
  \begin{itemize}
    \item
      We propose to fully consider the spatial-temporal context for periodic interest learning. Specifically, periodicities of daily, hourly, areal, and hourly-areal granularities are constructed.
    \item
      We propose two levels of attention. Intra-level attention learns long-term interest from the user's daily pattern. Inter-level attention integrates long- and short-term interests to obtain a better overall representation. 
  \end{itemize}

\section{Related Work}
  In this section, we briefly review some recent studies for the POI recommendation and summarize the differences. 

  Combining long- and short-term interests attracts much attention recently and it achieves state-of-the-art. DRCF \cite{manotumruksa2017deep} captures long-term interest by collaborative filtering and adds short-term interest by a recurrent neural network (RNN). STGN \cite{zhao2019go} modifies LSTM by devising spatial-temporal gates to enhance interest. DeepMove
  \cite{feng2018deepmove} applies an RNN-based method to capture transitional regularities and designs a historical attention module to exploit mobility periodicity by user status. LSPL \cite{wu2019long} and PLSPL \cite{wu2020personalized} use attention layer to obtain long-term interest and use LSTM to model recent sequential behaviors for short-term interest. LSTPM \cite{sun2020go} uses all trajectories with various techniques to obtain long-term interest, such as a time-weighted operation of periodicity, an LSTM and a geo-dilated LSTM are used to capture short-term interest. 

  The differences between previous work and our study are obvious. 
  Previous work usually captures periodicity from limited granularity, such as time slot. We fully utilize spatial-temporal context and devise effective attention operations to automatically combine various periodic interests.


\section{Methodology}
\begin{figure}[htpb]
    \centering
    \setlength{\belowcaptionskip}{-15pt}
    \includegraphics[width=\linewidth]{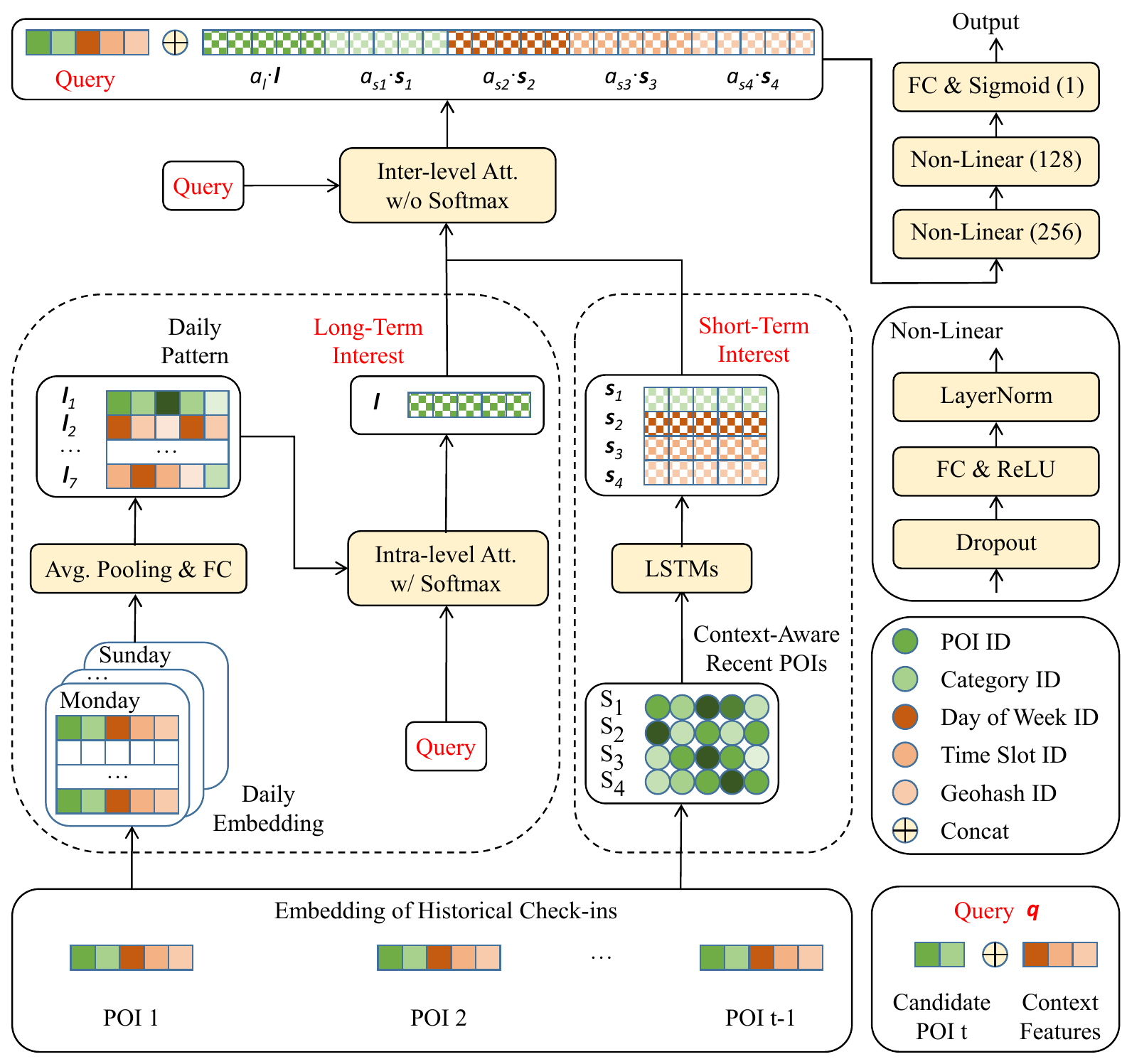}
    \caption{The overall architecture of ST-PIL.} 
    \label{fig:model}
  \end{figure}

  In this section, we introduce the details of our ST-PIL. 
  Fig. \ref{fig:model} shows the architecture, including an embedding layer, long- and short-term modules, inter-level attention and a prediction layer. Various context-aware short-term sequences are illustrated in Fig. \ref{fig:short_seqs}.

\subsection{Embedding Layer}
  First of all, we introduce rich features. Let $U=\{u_1, u_2, ..., u_{|U|}\}$, $P=\{p_1, p_2, ..., p_{|P|}\}$, and $C=\{c_1, c_2, ..., c_{|C|}\}$ denote the set of users, POIs, and categories, respectively, where each POI has a category $c \in C$. Next is spatial-temporal information. As each POI has a geocoding of latitude and longitude, we transform it into the geohash-integer version. We use geohash-5 $G=\{g_1, g_2, ..., g_{|G|}\}$ to express spatial information as areas, and each POI only belongs to one geohash-5. Spatial features are represented by day of week and time slot, where we have 7 days in a week $W=\{w_1, ..., w_7\}$ and 24 slots of hours in a day $M=\{m_1, ..., m_{24}\}$. Then, a user's check-in at $t$-th time step is $h^t=(p^t, c^t, w^t, m^t, g^t)$.
  In our work, each feature has a unique ID and is represented by a learned embedding.

  Next is the problem formulation. Given a user's all historical check-ins $H=\{h^1, h^2, ..., h^{t-1}\}$ as well as his/her current time and area (mapped as $(w^t, m^t, g^t)$), our goal is to select top-$k$ POIs that the user would check-in. For example, we now know that a user is near a certain mall, we need to predict which POI he/she will go to.

\subsection{Long-Term Module}
  In the long-term module, we use temporal context called day of week to find user's daily pattern and use intra-level attention to establish long-term interest.

  Daily pattern aims to acquire user's daily periodicity, from Monday to Sunday.
  Firstly, we construct daily mask sequences with the same length as the user check-in histories and divide all check-ins into different days to form daily embedding. 
  For example, we use a sequence $d = [1, 0, 1, ..., 0]$ to indicate the user performed check-ins on Monday, where symbols '1' and '0' refer to whether or not a check-in occurs on that day. Through the element-wise product between the mask $d$ and user's all historical check-ins, the check-ins on Monday are taken out separately.
  Next, through average pooling and a shared fully connected layer on each daily embedding, we end up with seven daily patterns, denoted as $\bm{L}=[\bm{l}_{1};\bm{l}_{2};...;\bm{l}_{7}]$. Specifically, we calculate the cumulative sums of a daily embedding and a mask and then divide them to achieve average pooling.

  \begin{figure*}[tb]
    \centering
    \setlength{\belowcaptionskip}{-15pt}
    \includegraphics[width=0.85\linewidth]{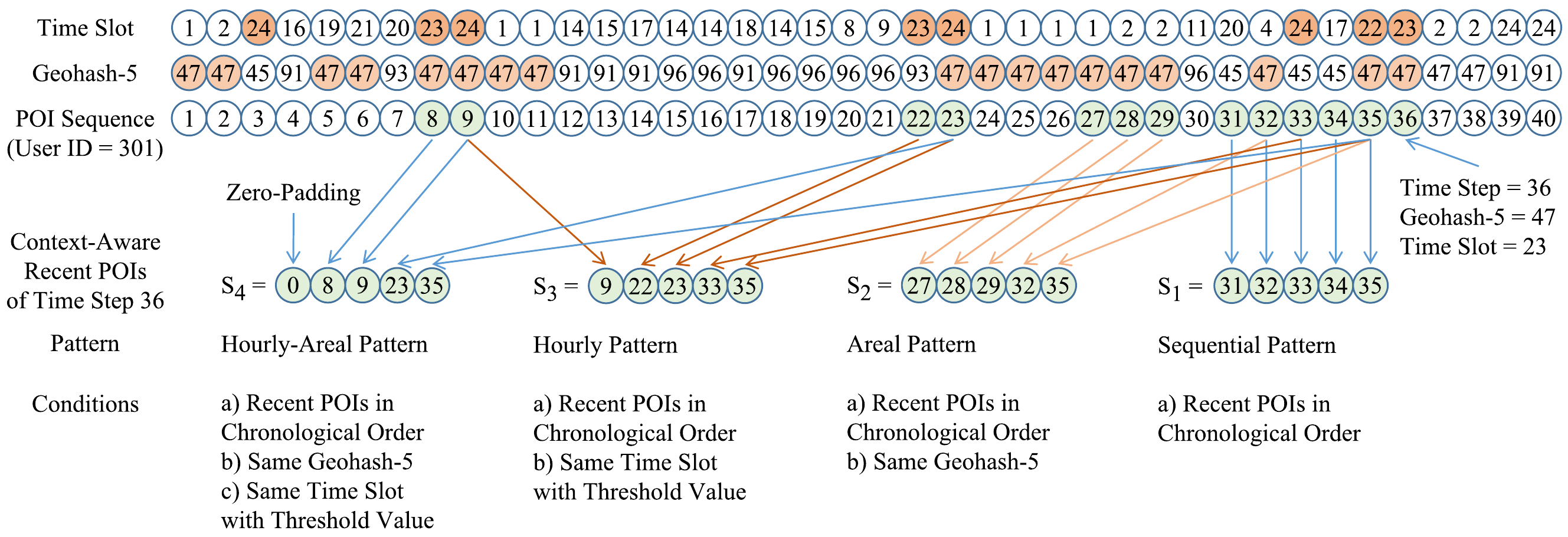}
    \caption{Illustration of obtaining various short-term sequences for a certain time step by spatial-temporal context.}
    \label{fig:short_seqs}
  \end{figure*}

  Long-term interest can summarize important information from the daily pattern with intra-level attention. We first create a query $\bm{q}$ for attention. 
  A query at $t$-th time step is built by the concatenation of a candidate POI and the already known context ($w^t, m^t, g^t$):
  \begin{equation}  \label{eq_query}
    \bm{q}^t = [\bm{p}^t, \bm{c}^t, \bm{w}^t, \bm{m}^t, \bm{g}^t],
  \end{equation}
  where every POI can be a candidate but only one is positive, and each feature ID is embedded by a dense vector. 
  Then, we apply the widely-used Bahdanau attention\footnote{\url{https://www.tensorflow.org/tutorials/text/nmt_with_attention}} \cite{bahdanau2015neural} to aggregate daily pattern $\bm{L}^t$:
  \begin{equation}    \label{eq_att_qk_long}
    \begin{split}
      e_j^t &= \bm{v}_e^{\rm{T}} {\rm{tanh}} (\bm{V}_1 \bm{q}^t + \bm{V}_2 \bm{l}_j^t), \\
      \bm{e}^t &= {\rm{softmax}} (\{e_1^t, e_2^t, ..., e_7^t\}), \\
      \bm{l}^t &= \sum_{j=1}^7 e_j^t \bm{l}_j^t,
    \end{split}
  \end{equation}
  where $\bm{v}_e, \bm{V}_1$ and $\bm{V}_2$ are weight matrices, each weight $e_j^t$  is normalized by $softmax$ and $\bm{l}^t$ is the $t$-th long-term interest.

\subsection{Short-Term Module}
  The short-term module plays an important role to capture the fundamental sequential pattern and context-aware patterns, e.g., hourly, areal, and hourly-areal periodicities. 

  Fig. \ref{fig:short_seqs} illustrates the rule conditions of obtaining short-term POI sequences $S^t=\{S_1^t, S_2^t, S_3^t, S_4^t\}$ from the user's all historical check-ins. In detail, at current time step $t=36$, we have spatial-temporal context called time slot $m^t=23$ and area $g^t=47$. Correspondingly, $S_1$ uses recent POIs to capture the sequential transition regularity, $S_2$ and $S_3$ finds POIs that the user has checked-in before in the current area $g^t=47$ and in similar time slots $\{22, 23, 24\}$ respectively, and $S_4$ is the insection of $S_2$ and $S_3$. 

  For each short-term sequence at $t$-th time step $S_k^t$, we apply a LSTM \cite{hochreiter1997long} to obtain its interest:
  \begin{equation}  \label{eq_short_interest_one}
    \bm{s}_k^t = {\rm{LSTM}} (S_k^t),
  \end{equation}
  where $\bm{s}_k^t$ is a separate short-term interest and $k=\{1,2,3,4\}$. As each sequence implies a specific mobility pattern, we use four LSTMs instead of one to avoid collision.

\subsection{Inter-level Attention}
  \noindent
  In this section, we obtain the final interest representation through inter-level attention.

  At current time step, we have five interests $\{\bm{l}^t, \bm{s}_1^t, \bm{s}_2^t, \bm{s}_3^t, \bm{s}_4^t\}$, where $\bm{l}^t$ is the long-term interest and others are short-term interests. We still use Bahdanau attention to compute weights, e.g., 
  \begin{equation}    \label{eq_att_qk_short}
      a_l^t = \bm{v}_a^{\rm{T}} {\rm{tanh}} (\bm{V}_3 \bm{q}^t + \bm{V}_4 \bm{l}^t),
  \end{equation}
  where $\bm{v}_a, \bm{V}_3$ and $\bm{V}_4$ are new weight matrices, and $a_l^t$ is the attention weight for $\bm{l}^t$. Correspondingly, other weights are computed for the rest interests. Next, we concatenate all these interests:
  \begin{equation}    \label{eq_att_qk_total_concat}
      \bm{x}^t = [a_l^t \bm{l}^t, a_{s1}^t \bm{s}_1^t, a_{s2}^t \bm{s}_2^t, a_{s3}^t \bm{s}_3^t, a_{s4}^t \bm{s}_4^t],
  \end{equation}
  where $\bm{x}^t$ is the overall interest. 

  Different from intra-level attention in Section 3.2, inter-level attention has its characteristics.
  First, as the interests all come from historical check-ins, we add a dropout layer for query $\bm{q}^t$ and interests to avoid overfitting before computing attention weights. Second, as these interests capture different mobility patterns, we aggregate them by weighted concatenation to preserve unique features. Besides, we abandon the $softmax$ on the attention weights to improve the expression ability of important patterns. 

  \begin{table*}
    \small
    \setlength{\abovecaptionskip}{0pt}
    \setlength{\belowcaptionskip}{-10pt}
    \caption{Performance comparison on two datasets. The best value and the runner-up are boldfaced and underlined, respectively.}
    \begin{tabular}{ccrrrrrrrrrr}
      \toprule
      & & \multicolumn{5}{c}{NYC} & \multicolumn{5}{c}{TKY} \\
      \cmidrule(lr){3-7} \cmidrule(lr){8-12} 
      & & Acc@1 & Acc@5 & Acc@10 & MRR@5 & MRR@10 & Acc@1 & Acc@5 & Acc@10 & MRR@5 & MRR@10 \\
      \midrule

      \multirowcell{4}{Baselines}
      & LSTM      & 0.1816 & 0.3991 & 0.4663 & 0.2610 & 0.2701 & 0.1506 & 0.3553 & 0.4511 & 0.2256 & 0.2384 \\
      & STGN      & 0.2071 & 0.4051 & 0.4645 & 0.2829 & 0.2909 & \underline{0.2134} & \underline{0.4352} & 0.5161 & \underline{0.2956} & \underline{0.3066} \\
      & LSTPM     & 0.1846 & 0.3693 & 0.4484 & 0.2549 & 0.2660 & 0.1928 & 0.3889 & 0.4622 & 0.2647 & 0.2746 \\
      & PLSPL     & \underline{0.2188} & \underline{0.4367} & \underline{0.5162} & \underline{0.3000} & \underline{0.3109} & 0.2054 & 0.4252 & \underline{0.5225} & 0.2853 & 0.2983 \\
      \cmidrule(lr){1-12}

      \multirowcell{4}{\textbf{Ours}}
      & $\textbf{ST-PIL}_{\rm{L}}$  & 0.3135 & 0.5543 & 0.6243 & 0.4055 & 0.4151 & 0.2842 & 0.5340 & 0.6287 & 0.3773 & 0.3900 \\
      & $\textbf{ST-PIL}_{\rm{S}}$  & 0.3790 & 0.5798 & 0.6363 & 0.4543 & 0.4619 & 0.3464 & 0.5872 & 0.6708 & 0.4386 & 0.4498 \\
      & \textbf{ST-PIL}             & \textbf{0.3807} & \textbf{0.5850} & \textbf{0.6454} & \textbf{0.4584} & \textbf{0.4666} & \textbf{0.3523} & \textbf{0.5963} & \textbf{0.6772} & \textbf{0.4455} & \textbf{0.4564}  \\
      & improvement (\%)              & +74.00 & +33.96 & +25.03 & +52.80 & +50.08 & +71.52 & +37.02 & +29.61 & +50.70 & +48.86 \\
      \bottomrule
    \end{tabular}
    \label{tab:performance_total}
  \end{table*}

\subsection{Prediction Layer}
  We compute the probability of a candidate POI based on the spatial-temporal context in the prediction layer, consisting of an input, a multilayer perceptron (MLP), and a loss function.
  The input for MLP is a concatenation of the query $\bm{q}^t$ and the overall interest $\bm{x}^t$:
  \begin{equation}    \label{eq_input}
    input^t = [\bm{q}^t, \bm{x}^t].
  \end{equation}
  The following process of MLP is illustrated in Fig. \ref{fig:model}. 
  Finally, we use the popular cross-entropy as our loss:
  \begin{equation}    \label{eq_log_loss}
    \mathcal{L} = - \sum_{i=1}^N y_i \cdot {\rm{log}}(p(y_i)) + (1-y_i) \cdot {\rm{log}} (1-p(y_i)), 
  \end{equation}
  where $N$ is the number of all candidates, including positives and negatives, $y_i \in \{0, 1\}$ and $p(y_i)$ are label and probability for the candidate respectively.

\section{Experiments}
\subsection{Experimental Settings}
  \noindent
  \textbf{Datasets.}  
  We choose two public real-world datasets collected from New York city (NYC) and Tokyo (TKY)\footnote{\url{http://www-public.it-sudparis.eu/~zhang_da/pub/dataset_tsmc2014.zip}} \cite{yang2014modeling}. Each check-in contains a user ID, POI ID, category ID, latitude, longitude, and UTC. 
  We delete POIs that are checked by less than 5 times, and we only keep the most recent 500 check-ins for each user.  
  Following the previous work \cite{manotumruksa2017deep}, we adopt the leave-one-out evaluation methodology. We treat each user's last POI as ground truth, compute probabilities for all POIs during the test and rank them. 

  \noindent
  \textbf{Baselines.}
  We compare our ST-PIL with several baselines, which are good at capturing long- and short-term interests. 
  \textbf{LSTM} \cite{hochreiter1997long} is a widely-used method for sequential modeling. 
  \textbf{STGN} \cite{zhao2019go} controls both interests by devising spatial-temporal gates. 
  \textbf{LSTPM} \cite{sun2020go} designs a non-local network for long-term interest and a geo-dilated LSTM for short-term interest.
  \textbf{PLSPL} \cite{wu2020personalized} uses user-based attention and LSTMs for two interests, respectively. And it adopts another user-based attention to obtain overall predicted probability.

  \noindent
  \textbf{Parameters and Evaluations.} 
  The dimensions of POI, category, day of week, time slot, and geohash-5 are 64, 32, 8, 16, and 32, respectively. We make these dimensions approximately proportional to the logarithm of the number of their unique ID values \cite{covington2016deep}. The number of negative candidates is 20. The dimension of each interest is 64. We grid search the dropout rate from $\{0.0, 0.1, 0.2\}$. Correspondingly, the best rates of inter-level attention are 0.0 and 0.1 on NYC and TKY, respectively, and the best rate of MLP is 0.1. We apply Acc@k \cite{xie2016learning} and MRR@k to evaluate methods. Our model has two variants, called ST-$\rm{PIL_L}$ and ST-$\rm{PIL_S}$ which only have the long-term module and short-term module, respectively.

\subsection{Results and Discussions}
  \noindent
  \textbf{Performance.}
  The results are listed in Table \ref{tab:performance_total} and our ST-PIL significantly outperforms the others. 
  Given the comparison with baselines, our ST-PIL achieves a great improvement on the two datasets. Specifically, for Acc@1, ST-PIL outperforms the runner-up by about 70\%. For Acc@5 and Acc@10, our model improves by about 30\%. The overall improvements of MRR@5 and MRR@10 are about 50\%. 
  In light of the model variant, ST-$\rm{PIL_S}$ outperforms ST-$\rm{PIL_L}$ and ST-PIL is the best, which clearly shows the advantage of our inter-level attention in integrating multiple interests.

\begin{figure}[h]
    \centering
    \setlength{\belowcaptionskip}{-15pt}
    \includegraphics[width=0.9\linewidth]{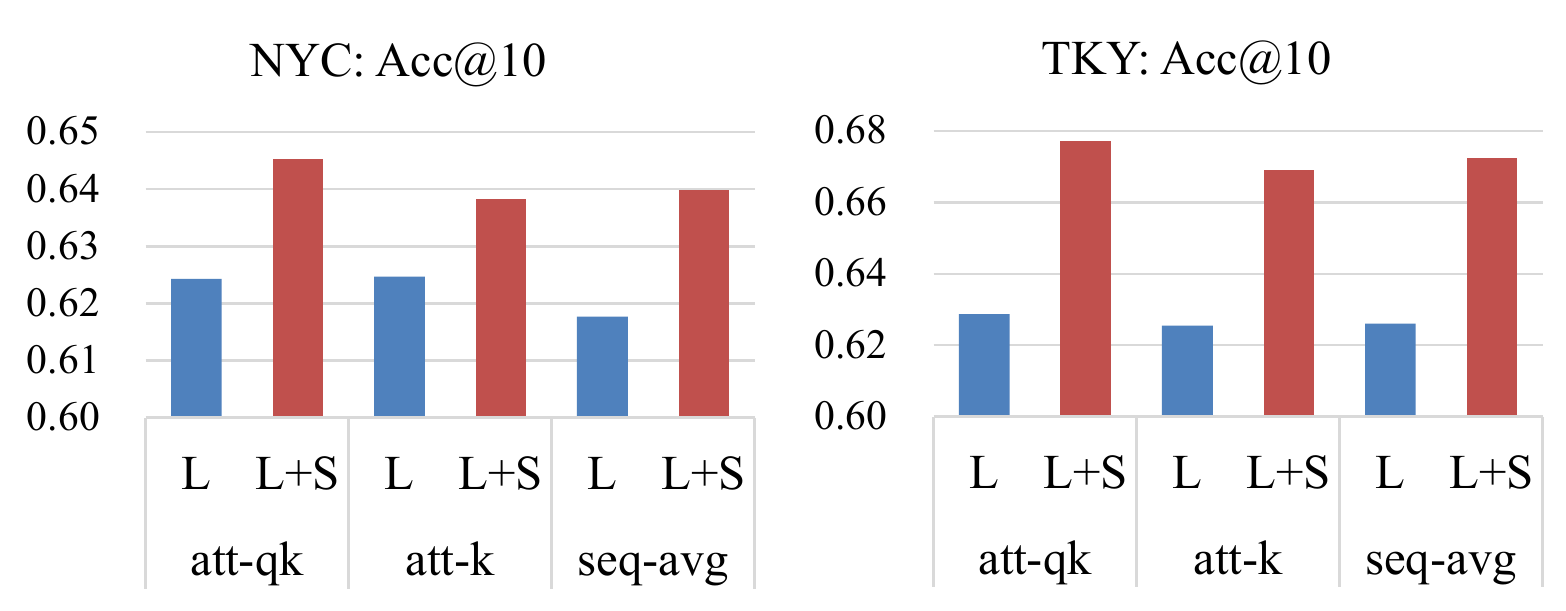}
    \caption{Examine different settings for long-term module.}
    \label{fig:exp_Long}
  \end{figure}

  \noindent
  \textbf{Settings of Long-Term Module.}
  We set up three settings, called att-qk, att-k, and seq-avg, to examine how to deal with all historical check-ins. The att-qk refers to our long-term module. The att-k does not use the query in intra-level attention. The seq-avg averages the entire POI sequence. The comparison is illustrated in Fig. \ref{fig:exp_Long}, where symbols 'L' and 'L+S' refer to ST-$\rm{PIL_L}$ and ST-PIL, respectively. 
  Obviously, the setting att-qk is best. In summary, capturing daily periodicity with a query can enhance the expressions of long-term interest and overall interest.

  \begin{figure}[h]
    \centering
    \setlength{\belowcaptionskip}{-15pt}
    \includegraphics[width=0.9\linewidth]{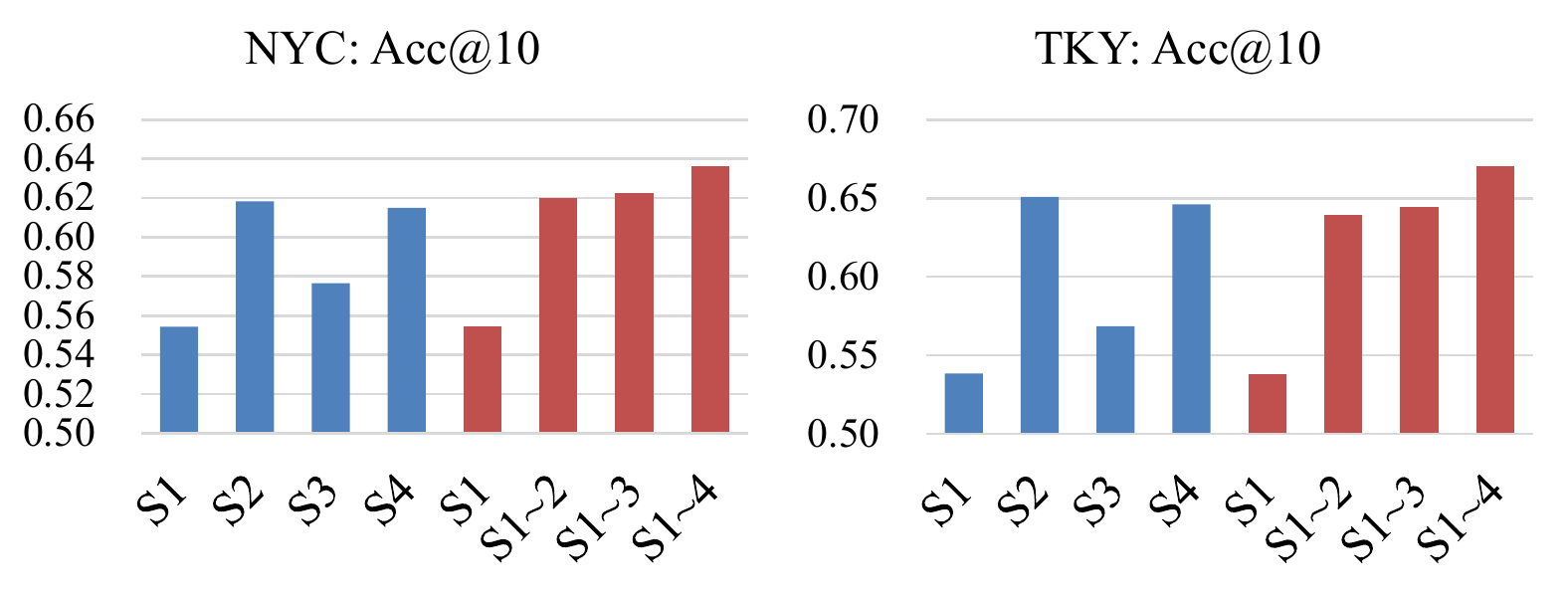}
    \caption{Ablation study of short-term module.}
    \label{fig:exp_Short}
  \end{figure}

  \noindent
  \textbf{Ablation Study of Short-Term Module.}
  We compare four short-term interests to explore different mobility patterns. Results are illustrated in Fig. \ref{fig:exp_Short} by blue and red colors. 
  For the first discussion, the four interests perform differently. The performance of $S_2$ and $S_4$ are better than the other two. This means the spatial and spatial-temporal periodicities are more influential than sequential effect and temporal periodicity. Perhaps when a user chooses a POI in a certain area, the POIs he/she has visited before in this area have a greater impact. 
  For the second discussion, the overall performance of ST-$\rm{PIL_S}$ is better than any short-term interest. Besides, when we gradually add four short-term interests, the overall performance is getting better and better. This comparison also verifies the effectiveness of our inter-level attention.

\section{Conclusion and Future Work}
  In this work, we propose a spatial-temporal periodic interest learning network (ST-PIL) with long- and short-term modules for the next POI recommendation. 
  In the long-term module, we explore the daily pattern and capture long-term interest by intra-level attention. In the short-term module, we extract various short-term sequences to model sequential, spatial, temporal, and spatial-temporal patterns. Inter-level attention can effectively integrate different interests. The experiment demonstrates the effectiveness of our ST-PIL. 
  In the future, we would design a more automated modeling methodology to replace the pre-operations of constructing different sequences.


\bibliographystyle{ACM-Reference-Format}


\end{document}